\documentclass{article}
\usepackage[a4paper, total={6in, 9in}]{geometry}
\usepackage[utf8]{inputenc}
\usepackage{float}
\usepackage{amsmath}
\usepackage{amsfonts}
\newtheorem{definition}{Definition}

\title{Identification of Anomalous Geospatial Trajectories via Persistent Homology}
\author{
  Evans-Lee, Kyle\\
  Systems and Technology Research\\
  Arlington, VA \\
  \texttt{kyle.evans-lee@str.us}
  \and
  Lamb, Kevin\\
  Systems and Technology Research\\
  Woburn, MA\\
  \texttt{kevin.lamb@str.us}
}
\date{August 2024}

\usepackage[numbers]{natbib}
\usepackage{graphicx}

\begin{document}

\maketitle

\begin{abstract}
We present a novel method for analyzing geospatial trajectory data using topological data analysis (TDA) to identify a specific class of anomalies, commonly referred to as crop circles, in AIS data.
This approach is the first of its kind to be applied to spatiotemporal data.
By embedding $2+1$-dimensional spatiotemporal data into $\mathbb{R}^3$, we utilize persistent homology to detect loops within the trajectories in $\mathbb{R}^2$.
Our research reveals that, under normal conditions, trajectory data embedded in $\mathbb{R}^3$ over time do not form loops.
Consequently, we can effectively identify anomalies characterized by the presence of loops within the trajectories.
This method is robust and capable of detecting loops that are invariant to small perturbations, variations in geometric shape, and local coordinate projections.
Additionally, our approach provides a novel perspective on anomaly detection, offering enhanced sensitivity and specificity in identifying atypical patterns in geospatial data.
This approach has significant implications for various applications, including maritime navigation, environmental monitoring, and surveillance.
\end{abstract}


\section{Introduction}
\begin{figure}[H]
    \centering
    \includegraphics[width=.49\linewidth]{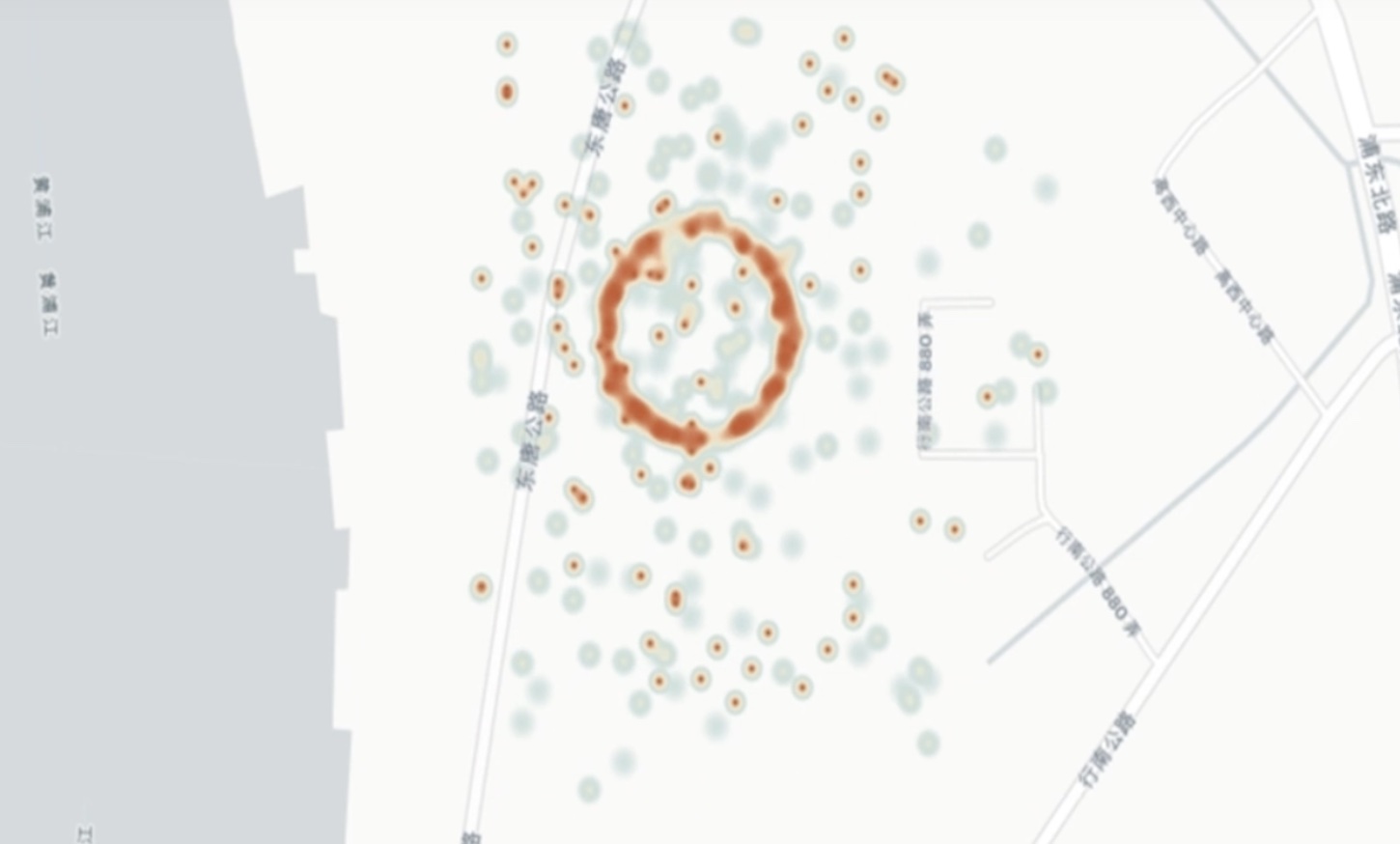}
    \includegraphics[width=.42\linewidth]{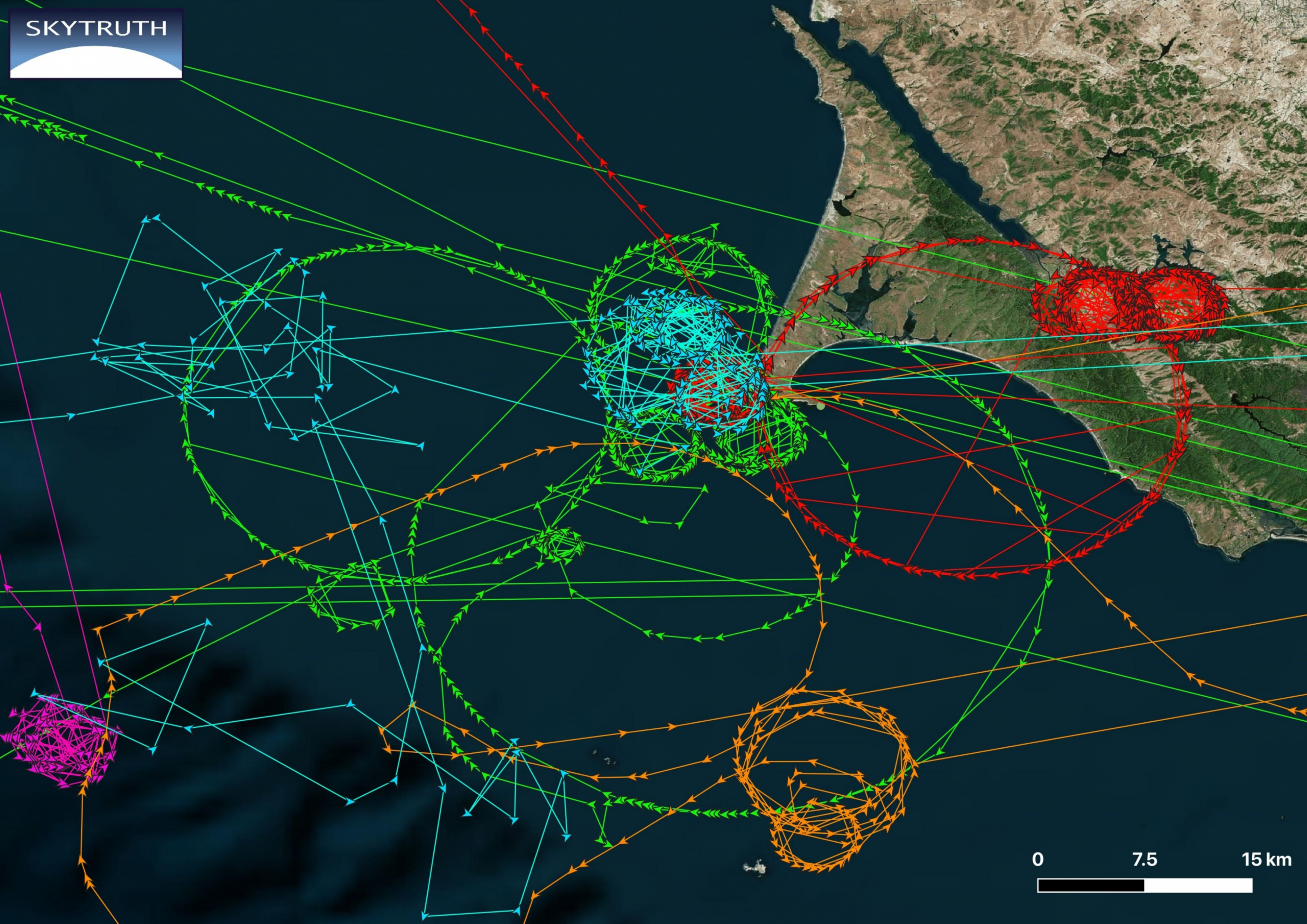}
    \caption{(Left: Example of crop circles found in Shanghai \protect\cite{Harris}. Right: Example of continuous circles found near San Francisco, a region with known anomalies \protect\cite{skytruth}.}
    \label{fig:track_augmentation}
\end{figure}

A class of anomalies referred to as \textit{crop circles} are increasingly being found in maritime geospatial data \cite{Harris}.
These circular anomalies (see Fig. \ref{fig:track_augmentation}) can be composed of multiple selectors; that is, tracks containing multiple crop circles.
For this work, however, we will be identifying anomalies at an individual selector level.
These anomalies have been found all over the world; their regularity of occurrence is increasing; and the motivation and reason for their existence is unclear \cite{Harris}\cite{spirent}\cite{wpsanet}.
Crop circles are often difficult to identify as they are not necessarily perfectly circular, possibly due to spherical projections, track perturbations, or movement of center/radius. 
Here we present a scalable approach to identifying these anomalies using techniques from topological data analysis (TDA) and algebraic topology.
This is accomplished by calculating topological features of tracks via persistent homology.
These calculations are scalable since the computational complexity of persistence calculations are small on tracks and CPU-parallelizable across different selectors.

\begin{figure}[H]
    \centering
    \includegraphics[width=.4\linewidth]{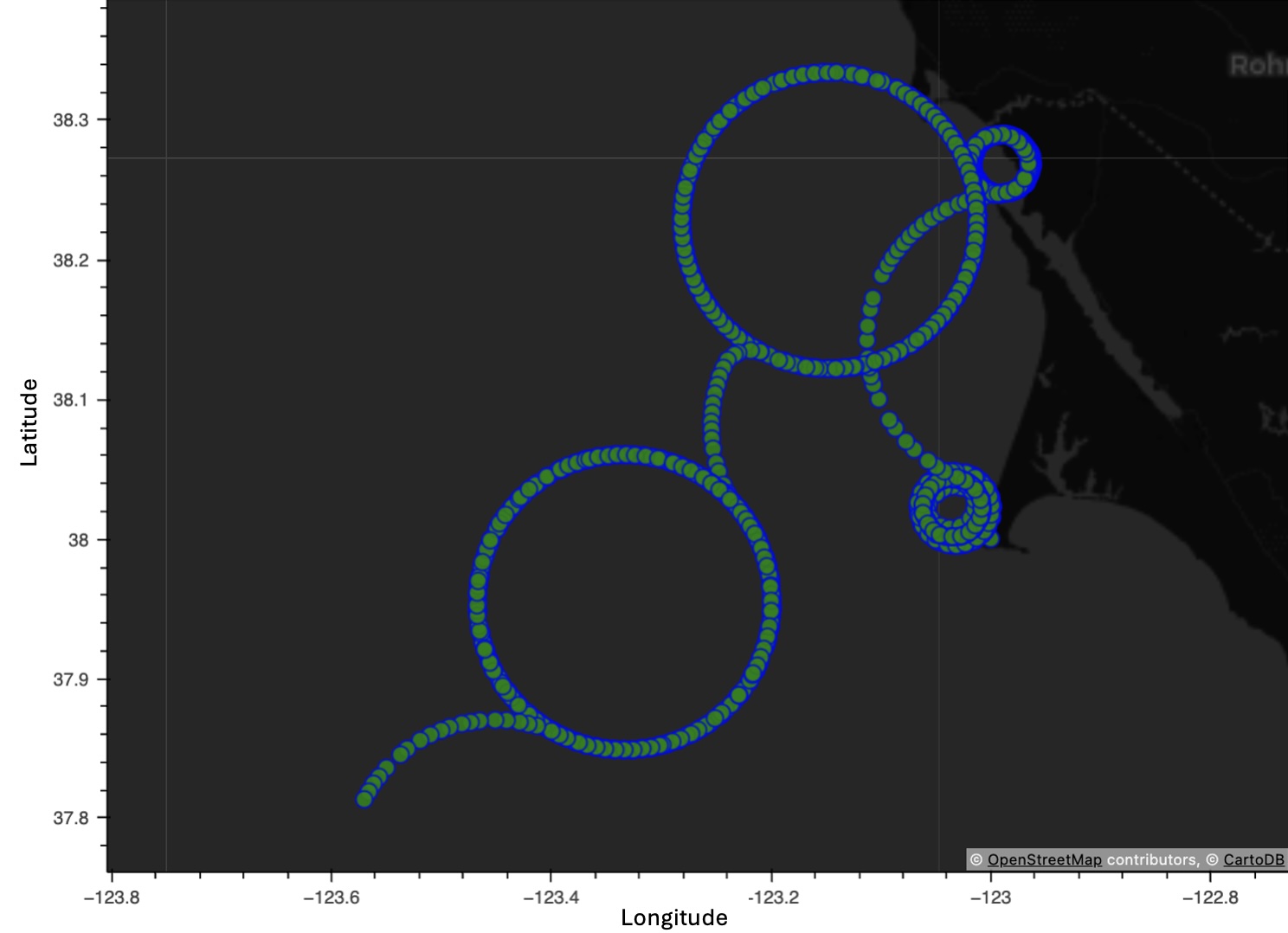}
    \caption{Cargo ship identified using features derived from persistent homology of AIS trajectory data near Point Reyes}
    \label{fig:crop_simple}
\end{figure}

Fig. \ref{fig:track_augmentation} shows two examples of  circles in AIS data from 2019, one near Shanghai and one near San Francisco.
Both examples are known to be anomalous from satellite imagery of the vessel \cite{Harris}\cite{skytruth}. 
The second image shows the region we will be using to identify new anomalies using persistent homology.
We were able to apply persistent homology to AIS trajectory data on one year's worth of data to identify ships showing anomalous geospatial coordinates.
Fig. \ref{fig:crop_simple} shows an outlier detected using features derived via our TDA method.

Fig. \ref{fig:crop_simple} shows a previously unknown example of a crop circle off Point Reyes.
It was detected using features derived via TDA methods.
In particular, this highlights the ability of topological features to identify certain classes of anomalous geospatial trajectories.

\subsection{Related Work and Novel Contributions}
Anomaly detection, topological data analysis, and geospatial data analysis are well-researched topics.
Our novel approach focuses on the application of TDA to analyze track-level spatiotemporal data, presenting a unique method for effectively identifying anomalies.

\subsubsection{Anomaly Detection Within Geospatial Data}
Significant work has been done on the identification of anomalies within geospatial data developing techniques to find unusual velocities, shapes, and locations.
Large datasets such as Automatic Identification System (AIS) (provided when a ship transmits its location) or Automatic Dependent Surveillance–Broadcast (ADS–B) (provided when an aircraft transmits its positioning source) are known to have unusual records \cite{gnss}\cite{Harris}.

\subsubsection{TDA in Geospatial Data}
TDA has been applied to geospatial data in the past; however, most approaches look at the datasets as a whole to extract information from the shape of the entire dataset.
This has included the extraction of meaningful movement patterns in AIS data \cite{aismovement} or in voting patterns across states \cite{vote}.

\subsection{Our Contributions}
Current approaches for using TDA on spatiotemporal data utilize population-level analysis on the density manifold \cite{TDAgeospatial, vote, CityGML}.
These approaches garner insights into the topology of large data sets but can be computationally expensive to compute.

Our approach is fundamentally different in two ways.
First, we look at topological features at a selector level and, hence, these features can be easily computed in parallel.
Second, we notice that spatiotemporal trajectories should not have any meaningful loops in three dimensions when considered with time.
Thus, the existence of significant generators in $1^{\text{st}}$ homology can be thought of as an obstruction to a track being ``normal''.
Specifically, trajectories embedded in $\mathbb{R}^3$ having a long-lived generator in its $1^{\text{st}}$ persistent homology must be anomalous.

\section{Background and Detection Strategy}
This section begins by reviewing the underlying mathematical techniques used in our approach.
We then explain how to apply these techniques to AIS data.

\subsection{Geospatial Data}
A single track in geospatial datasets such as AIS data consists of time series of ordered triples $\{(x_i,y_i,t_i)\}$.
The first two coordinates indicate the position of a ship, and the third indicates the time at which the measurement was taken along with a unique selector value $s$.
In general, the spatial coordinates $x_i$ and $y_i$ are all measured in decimal degrees, and the times $t_i$ are measured in seconds.
We analyze this data by first transforming all data associated with a selector into a point cloud in a higher-dimensional Euclidean space $\mathbb{R}^n$.
It is within this point cloud that we search for patterns indicating the presence of crop circles in the track.

\subsection{Standard Approaches for Analyzing Geospatial Data}
High-dimensional point clouds of data can be difficult to analyze due to our lack of ability to adequately visualize the complexity of any underlying patterns they follow.
Dimension reduction is a common strategy in data analysis that addresses this concern.
In principle, dimension reduction identifies subspaces of the dataset where patterns and associations within the data are more easily visualized and analyzed.
The process is commonly divided into two steps: (1) matrix factorization and (2) construction of neighborhood graphs.

Algorithms like principal component analysis (PCA) \cite{PCA} can accomplish step (1) and use the results to identify macroscopic trends in the data.
However, this approach and others like it largely fail to identify nonlinear subspaces along which the crop circles' point-cloud features will lie.
Nonlinear methods like Sammon mapping \cite{Sammon} and kernel PCA \cite{kPCA} adapt to nonlinearities in the data trends, but they require solving a nonlinear optimization problem or training a classifier.

Step (2) can be addressed with techniques like $t$-distributed stochastic neighbor embedding (t-SNE) \cite{tSNE} or manifold learning \cite{MnfdLearn}.
However, the results of these methods are sensitive to the chosen probabilistic model for data association, or they must assume that the data is sufficiently dense to adequately represent the underlying manifold.

Because we are searching a point cloud for circles, we might consider using a circle detection algorithm.
The Hough transform \cite{gonzalez2018digital} is a common tool used in computer vision for circle detection.
While this may be a valid approach when the anomalies are geometrically perfect circles, it will fail in various ways if the circles are deformed, as is commonly the case for circular routes viewed off-center.

One of the major advantages of TDA is its robustness against the perturbation of geometric features.
It focuses instead on the underlying topological features of the dataset; that is, features that are invariant under continuous deformation of shape.

\subsection{TDA for Geospatial Data}
TDA is a mathematical and computational approach to data analysis that employs principles from algebraic topology to study the underlying structure and shape of a data set \cite{TDA}.
TDA focuses on capturing and quantifying the topological features within the dataset such as connected components, loops, and higher-dimensional voids.
This affords a more robust and holistic understanding of its geometric properties, since geometry more rigidly quantifies properties of topological spaces.
Additionally, it enables the identification of geometric features in the reduced data that associate directly with important descriptive features of the original data \cite{Mapper,CancerTDA,NatImgs}.

\subsection{Takens' Embedding}
The construction outlined in this section is a standard approach for embedding scalar and vector time series data as a point cloud in a high-dimensional Euclidean space.
The details beyond the next paragraph may be omitted on a first reading without any loss of general context in the overall approach to the detection of crop circles in geospatial data.

AIS track data is arguably already presented as a point cloud in two senses: (1) as a collection of  two-dimensional points on the surface of the earth and (2) as a collection of points $\{(x_i,y_i,t_i)\}\subset\mathbb{R}^3$.
Point clouds generally do not contain any sense of sequence, but the temporal nature of AIS track data gives us additional context for data analysis that we can leverage.
In the first sense, the data are sequential but lose the kinematic information they contain.
Hence, kinematically inconsistent spatial associations of the track data would pass unnoticed without the context provided by the timestamp data.
Consequently, some circular behavior may be observed but would not constitute anomalous behavior within the track.
In the second sense, circular behavior cannot be observed since recurrent spatial locations in a track occur at different times.
Therefore, the track data must be transformed in such a way that anomalous spatial behavior can be objectively identified within the context of the kinematics provided by the temporal data.

The \textit{Takens' embedding}, or \textit{time-delay embedding}, is a signal processing data transformation explicitly design to detect periodic behavior in time series data (although it is worth noting that the original application for Takens' work \cite{Takens} was to the study of strange attractors in dynamical systems).
It is a sliding-window technique that transforms a time series into a point cloud in a high-dimensional Euclidean space $\mathbb{R}^L$.
To do this, consider a time series $\{x_i\}_{i=0}^N$ of real numbers, and fix a positive integer $L\leq N-1$.
For each $i=0,1,\ldots,N-(L-1)$, define the point $y_i = (x_i,x_{i+1},\ldots,x_{i+(L-1)})\in\mathbb{R}^L$.
The sequence $\{y_i\}_{i=0}^{N-(L-1)}$ is called the \textit{dimension-$L$ Takens' embedding} of the time series $\{x_i\}_{i=0}^N$.
The integer $L$ is called the \textit{window length}.
It is conflated with the dimension of the embedding and so is often called the \textit{dimension}.

If the time series $x$ is given as a function of time, then there are additional parameters that can be tuned that produce interesting downsamplings of $x$.
Writing $\tilde{x} = (x(t_i),x(t_i+\tau),x(t_i+2\tau),\ldots,x(t_i+(L-1)\tau))\in\mathbb{R}^L$, the \textit{stride} of the Takens' embedding $\tilde{x}$ is the difference $t_{i+1}-t_i$ and the \textit{time delay} is the parameter $\tau>0$.

The Takens' embedding can also be applied to time series of vector data. This is accomplished by first passing the vector time series $\{(x_{i,1},x_{i,2},\ldots,x_{i,M})\}_{i=0}^N\subset\mathbb{R}^M$ to an auxiliary time series
\[
\{x_{0,1},x_{0,2},\ldots,x_{0,M},x_{1,1},x_{1,2},\ldots,x_{1,M},\ldots,x_{N,1},x_{N,2},\ldots,x_{N,M}\}.
\]
The dimension-$L$ Takens' embedding of $\{(x_{i,1},x_{i,2},\ldots,x_{i,M})\}_{i=0}^N$ is then defined to be the Takens' embedding of the auxiliary time series.
The dimension $L$ of the Takens' embedding of a vector time series is chosen to be some multiple of the dimension $M$ of the vectors to maintain the proper association between the original time series and its auxiliary time series.
This will help to avoid aliasing effects and erroneous associations across distinct components of the vectors.

Observe also that this embedding reproduces the original vector time series in $\mathbb{R}^M$ when the dimension $L$ and the stride of the embedding are both $M$ and its time delay $\tau=1$.
In AIS track data, the sliding window nature of this method incorporates and maintains the temporal relations between spatial coordinates $x_i, y_i$ in the time series.
For example, the consecutive points $(x_1,y_1,t_1)$ and $(x_2,y_2,t_2)$ in an AIS track will produce the points $(x_1,y_1,t_1)$, $(y_1,t_1,x_2)$, $(t_1,x_2,y_2)$, and $(x_2,y_2,t_1)$ in the Takens embedding we use for crop circle detection.

It is an immediate corollary from \cite{Takens} that the presence of recurrent trajectories (i.e., loops) in the dimension-$L$ Takens' embedding indicates that the signal is $L$-periodic.
This now gives us a methodology by which we are able to detect anomalous track data.
That is,
\begin{quote}
    a track will be considered anomalous if there exists a loop in the dimension-$L$ Takens' embedding for some dimension $L$.
\end{quote}
For the the calculations below we consider $L=3$; however, we wish to emphasize the generality of the construction and that further study is necessary to determine which other dimensions, if any other than $L=3$, will be useful in application to geospatial data analysis.

\subsection{Persistent Homology}
The primary tool in TDA is persistent homology, which tracks how topological features in the data persist over increasing spatial scales \cite{PersHom}.
Persistent homology is computed from a point cloud of data, resulting in a sequence of simplicial complexes (i.e., higher-dimensional generalizations of filled triangles).
From these simplicial complexes, a sequence $\{\left(b_0,b_1,\ldots,b_K\right)_n\}_{n=1}^N$ of ordered $(k+1)$-tuples of non-negative integers called \textit{Betti numbers} is produced.
The $k^\text{th}$ Betti number $b_k$ counts the number of $k$-dimensional voids in the point cloud.
These voids describe structural features of the point cloud like the numbers $b_0$, representing a point cloud with no voids (\textit{i.e.}, a cluster), and $b_1$, representing loops.
Not only, then, does persistent homology reduce the dimensionality of the original data set, but it differs from other nonlinear dimension reduction techniques by simultaneously analyzing the reduced data set and discovering finer and higher-dimensional features that exist within it.
This makes TDA a potentially powerful tool for the identification of anomalies in AIS track data.

\begin{figure}[H]
    \centering
    \includegraphics[width=\linewidth]{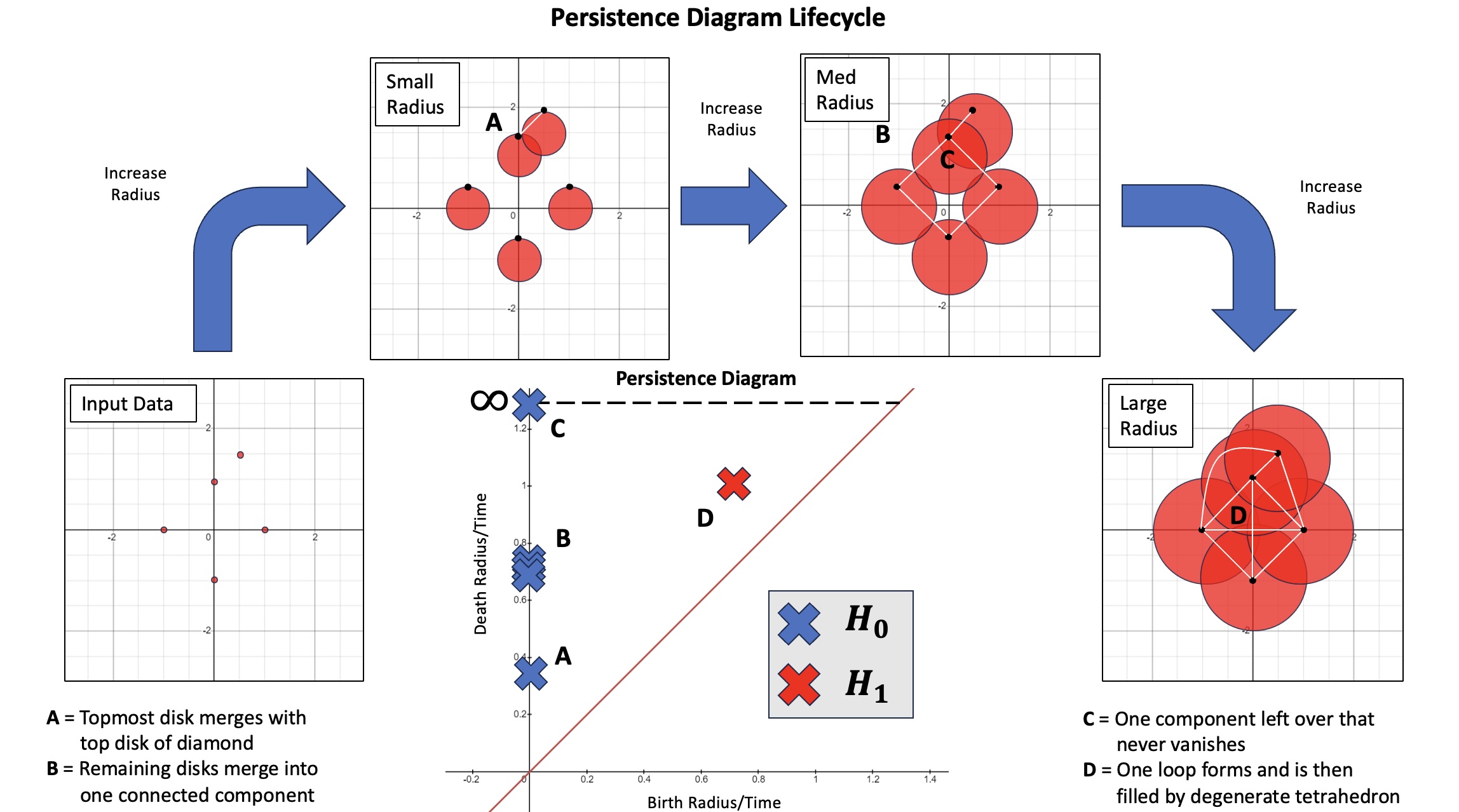}
    \caption{An example construction of a filtration and its associated persistence diagram. Disks (red) are inflated around five points (black) in the plane, and edges (white) are drawn between their centers according to the intersections of the disks to mark the Rips complexes $\Delta_r$ constructed at each stage. A persistence diagram notes the births and deaths of the topological features of $\Delta_r$ as the radius increases to infinity. Noteworthy events A, B, C, and D occur high above the diagonal in this graph. The stacked crosses in event B are located at the same point since they all occur simultaneously. Notionally, features that never die have a death time of $\infty$, as is the case with event C.}
    \label{fig:RipsComplexExample}
\end{figure}

To compute the persistent homology of a point cloud, we must first construct a sequence of simplicial complexes from the point cloud as follows.
We imagine gradually inflating a sphere of radius $r\geq0$ around each data point as in Fig. \ref{fig:RipsComplexExample}.
As $r$ increases, adjacent spheres will begin to intersect.
When $k$ many spheres intersect, we construct a k-simplex spanning their centers.
\begin{definition}
  A $ \textbf{k-simplex}$   is the k-dimensional polytope that is the convex hull of its $k + 1$ vertices defined as: 
  ${\displaystyle \left\{x\in \mathbb{R} ^{k}:x_{0}+\dots +x_{k-1}=1,x_{i}\geq 0{\text{ for }}i=0,\dots ,k-1\right\}.}$
  
\end{definition}
For low dimensions, a 1-simplex is a closed the unit interval, a 2-simplex is a filled triangle, and a 3-simplex is a filled tetrahedron.

For each $r\geq0$, there is a simplicial complex $\Delta_r$, called the \textit{Rips complex}, constructed in this way (allowing simplices to degenerate when their dimension exceeds that of the the Euclidean space containing the point cloud).
In the literature, the Rips complex is also called the \textit{Vietoris-Rips complex} or abbreviated as the \textit{VR complex}.
See Fig. \ref{fig:RipsComplexExample} for an example construction.

We note here that $\Delta_r$ is a subcomplex of $\Delta_R$ whenever $r < R$ in the Rips complex.
It is critical to clarify, however, that the topological features of $\Delta_r$ do not necessarily exist within $\Delta_R$.
Therefore, it is important to track how the topology of $\Delta_r$ changes as $r$ increases.
The sequence $\{\Delta_r\}_{r\geq0}$ is called a \textit{filtration} of the point cloud.

The term ``persistent'' in ``persistent homology'' refers to those topological features that remain from one Rips complex to the next in a filtration.
This is evident in the filtration's \textit{persistence diagram}.
Each point $(b,d)$ in this diagram represents a particular topological feature within the data that is first seen in $\Delta_b$ and disappears in $\Delta_d$.
The labeling identifies the dimension of that feature.
For example, a $0$-dimensional feature is a connected component of the data, and a $1$-dimensional feature is a loop in the data.
A point in the persistence diagram is often conflated with its associated topological feature in $\Delta_r$ for all $b\leq r\leq d$.
It is also worth noting that the parameter $r$ is often regarded as a notion of time rather than scale, and we choose to adopt this convention as well.
In this sense, the values $b$ and $d$ represent the times of \textit{birth} and \textit{death}, respectively, of topological features in the Rips complexes of the filtration.
Fig. \ref{fig:RipsComplexExample} contains the persistence diagram for the example construction.

For a feature $(b,d)$ in the persistence diagram, the difference $d-b$ is called called that feature’s \textit{lifetime}.
The guiding principle of persistent homology is that the longer a feature's lifetime, the more significant that feature is in its point cloud.
Contrapositively, features with short lifetimes are considered noise-like.
Noise-like features are indicated by points near the diagonal given by the relation $d=b$, while the significant features are typically located high above it.

It is impractical to compute a filtration using all possible real values of $r$.
The diameter of the convex hull of the point cloud provides an obvious upper bound beyond which no topological changes occur (no more associations between points can be made beyond this scale).
However, a discretization must be chosen carefully so as to (1) detect the changes in topology between the $\Delta_r$ complexes and (2) be fine enough to measure the persistent features, yet (3) be coarse enough to be computationally practical.
We have chosen a TDA software package, \ttfamily giotto-tda \normalfont \cite{giotto}, implemented in Python, that automatically computes (via a \textit{Ripser} C++ backend \cite{Bauer2021}) an efficient refinement of the filtration ensuring that all events are captured.

Our approach to detecting crop circles in AIS track data, then, is to
\begin{quote}
    compute a filtration of an appropriate Takens' embedding and search for long-lived $1$-dimensional features of the persistent homology.
\end{quote}
It is important to emphasize that the dimension of the embedding is not specified \textit{a priori} in this analysis.
A hyperparameter sweep for choosing this parameter in an optimal sense may be performed; but for the purposes of this work, we minimize the number of hyperparameter dependencies in this study by considering only Takens embeddings of dimension 3.
This is a reasonable choice that ensures each point in the track data has a direct representative in the Takens embedding.

\subsection{Metric}
The construction of a filtration for computing the persistent homology of a data set necessitates a well-defined metric on the data points.
This metric is essential for determining distances between points, which in turn dictate the formation of edges in the Rips complexes.
For trajectory data, which inherently involves both spatial (longitude, latitude) and temporal (timestamp) dimensions, the metric  must be defined on the space of $(\text{longitude},\text{latitude}, \text{timestamp})$ triples.
Ignoring the temporal dimension entirely would lose essential information about movement patterns and temporal evolution.

\subsubsection{Construction of Metric}
The suitable metric can be constructed by slightly modifying the standard Euclidean distance, accounting for the incommensurate units of time and distance.
The modification we make converts the temporal data into spatial data by multiplying it by an appropriate scaling factor.
We seek a velocity parameter $k>0$ (in units of km/hr) that will effectively equate $1\text{ hr}$ to $k \text{ km}$.
This is similar to the concept of reachability and defines a natural spatial scale for the data set.

For any choice of $k$, we define
\[
d(x,y) = [(x_0-x_1)^2+(y_0-y_1)^2+(k(t_0-t_1))^2]^{1/2},
\]
where $(x_i,y_i)$ are latitude-longitude coordinate pairs and $t_i$ are the times when those locations were measured.
Note that it is necessary to locally project latitude and longitude using the appropriate projections, such as those contained in the Spatial Reference System Identifier (SRID) registry, in order for the units of latitude and longitude to be in units of km and to be scaled properly.
Alternatively, we could use the Haversine (great circle) distance between longitudinal points (instead of $x_0-x_1)^2+(y_0-y_1)^2$)  which would yield a very similar metric. 
However, since homological calculations are robust to small perturbations of the shape and the sphere is locally Euclidean, this distinction isn't meaningful for our calculations.
It must also be acknowledged that some care must be taken to avoid discontinuities in lat-lon coordinates of significantly long tracks, but this does not affect the definition of the metric given above provided the tracks do not span all longitudes.

\subsubsection{Velocity Parameter Tuning Methods}
There are two ways to determine a suitable value of the parameter $k$: local and global.
A local parameter would mean that $k$ would be a function of latitude, longitude and, in principle, could also be a function of time.
A global parameter would be a constant $k$ that does not vary with respect to location or time.
The choice of method depends on the variation of the spatiotemporal scale of the data set as well as the features of interest.
In either case, hyperparameter tuning must be performed to find an optimal value of $k$, and we aim to choose the method that finds the value of $k$ maximizing the separation of persistent features between anomalous and non-anomalous tracks.

The first method for defining a metric assumes a local velocity, allowing the value of $k$ to vary across the spatial and temporal region of the data set.
A different value of $k$ could be chosen from location to location.
This amounts to choosing an open neighborhood around each point and defining a different velocity scale within each of those neighborhoods.

The second method assumes a global velocity, fixing a constant value of $k$ throughout the entire region.
A reasonable way to define a local velocity is to use the median velocity within that neighborhood.
This affords a quasi-metric $d$ on $\mathbb{R}^3$ as the distance between two points; that is, a function $d:\mathbb{R}^3\times\mathbb{R}^3\to\mathbb{R}$ that is positive-definite and satisfies the triangle inequality, but is not symmetric.
For the computations needed in persistent homology, we only need this quasi-metric as we pre-compute a pairwise distance table.
When computing persistent homology, two points for a given $\epsilon>0$ are identified when $\min\{d(a,b),d(b,a)\}<\epsilon$.
This can be thought of in terms of travel: the time needed to travel from point $A$ to point $B$ is not necessarily the same as from $B$ to $A$ (such is the case for two-way traffic; see section $5.7$).

This global approach proves to be sufficient for distinguishing crop circle features in the chosen data sets, but the local definition of $k$ could be a better choice for other applications or data sets spanning disparate scales (e.g., urban versus highway traffic).
As will be discussed in Section 3, a hyperparameter sweep for a global velocity scale shows an optimal value for $k$ is $10\text{ km/hr}$.




\subsection{Novelty of Approach}
\begin{figure}[H]
\centering
\includegraphics[width=.9\linewidth]{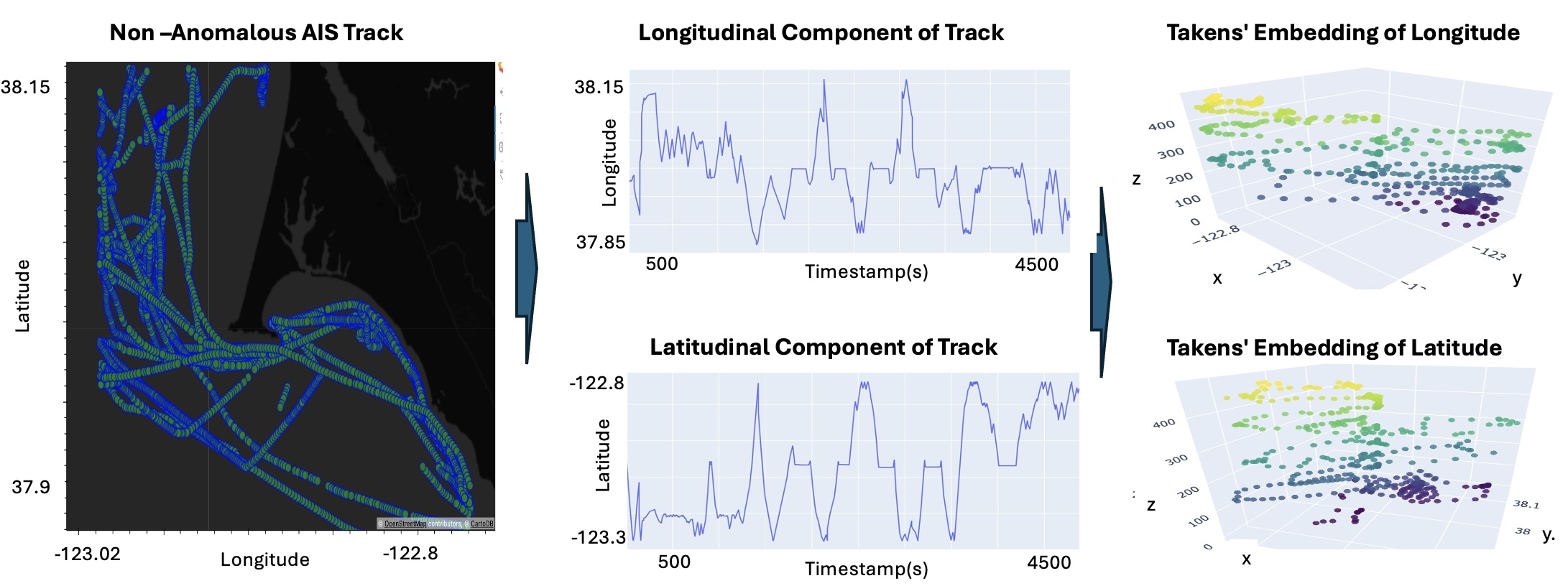}
\caption{Example of non-anomalous AIS track around Point Reyes 2020 (left). The individual latitudinal and longitudinal components of the track over time are shown (center) along with their dimension-3 Takens embeddings (right). Note that there are loop features present in the Takens embeddings of the individual latitude and longitude signals, but no anomalous behavior is observed in the track data.}
\label{fig:1dtak}
\end{figure}

TDA has been employed on geospatial data in the past, but there are several fundamental differences between that work and our approach.
First, we apply persistent homology calculations at the track level.
This makes the approach significantly more scalable and CPU-parallelizable, since individual tracks tend to be quite small.
Second, instead of using loops to identify features of the data, we instead view loops as an obstruction to a track to being normal.
In this manner, we are using persistent homology to perform outlier detection on the space of tracks, an approach distinct from other uses of persistent homology. 
Third, the use of TDA mitigates several problems associated with perturbations, local EPSG projections (as noted above), orientation of elliptical axis, and non-circular anomalies.
Finally, and perhaps most distinctly, we analyze the latitude and longitude signals jointly.
It is indeed possible to analyze these signals as independent time series, but consider the following example illustrated in Fig. \ref{fig:1dtak}.

Fig. \ref{fig:1dtak} shows how Takens' embedding treats typical latitudinal and longitiudinal data when these coordinates are analyzed separately.
Looping behavior in the embedding is not only present but expected in the dimension-3 Takens embedding.
Latitude and longitude values will naturally return often to the same value, but this behavior is not necessarily correlated with anomalous track data.
Although in principle there can be loops in the latitudinal or longitudinal signal, such constructions have never been seen and are extremely contrived.
It is only when considering  $(\text{latitude},\text{longitude},\text{timestamp})$ together that loops in the embedding become anomalous.
Again, the loops in the track data or a Takens embedding need not be perfect circles in order to be detected.

To summarize, our approach to finding crop circles in geospatial data is to
\begin{enumerate}
    \item
    embed individual trajectories into $\mathbb{R}^3$ via Takens' embedding in order to maintain spatiotemporal relationships,
    \item
    compute the pairwise distance between points in the trajectory using our modified Euclidean distance that accounts for unit ambiguities and natural scaling of the data,
    \item
    compute the persistence diagram for the $1^{\text{st}}$ homology where we would expect to find loop topology in the embedding, and
    \item
    extract features from that persistence diagram and identify outliers that may exist there.
\end{enumerate}

\section{Calibrating With GPS Data}
In the methodology laid out in the previous section, there is only one parameter that needs to be chosen empirically: the velocity parameter $k$ in the distance metric.
Due to its accessibility, abundance, and relative similarity to ship tracks in AIS, we have opted to use the GPS data from buses in Rio de Janeiro, Brazil, to identify an optimized value of $k$ that will adapt reasonably well to AIS track data.

Since this dataset does not contain the kind of anomaly we seek in AIS ship tracks, we need to reserve some subset of the bus data and modify it to contain perfect and perturbed crop circles.
Once this is done, we will perform a hyperparameter sweep over the various values of $k$ to find the one that maximizes the separation between real and augmented track data.

\subsection{Synthetic Data Generation}
We use the Rio Bus Dataset as our base truth data for tracks and determine the velocity scale parameter for the metric \cite{balteiro2023gps}.
The Rio Bus Dataset consists of GPS records for more than 9,000 buses in Rio de Janeiro across approximately 490 bus routes from September 26, 2013, to January 9, 2014. Each entry includes the bus ID, bus route ID, timestamp in UTC format, latitude, and longitude.
We segment each bus track into smaller tracks based on the assumption that, if it is at rest for more than 45 minutes with no update, it should be segmented into a new track.
All tracks in this dataset are assumed to not be anomalous.
We generate anomalous tracks by selecting 10\% of the data and augmenting the middle third of a track to be anomalous with kinematics generated using the exemplar from \cite{Harris}.


\begin{figure}[H]
    \centering
    \includegraphics[width=.4\linewidth]{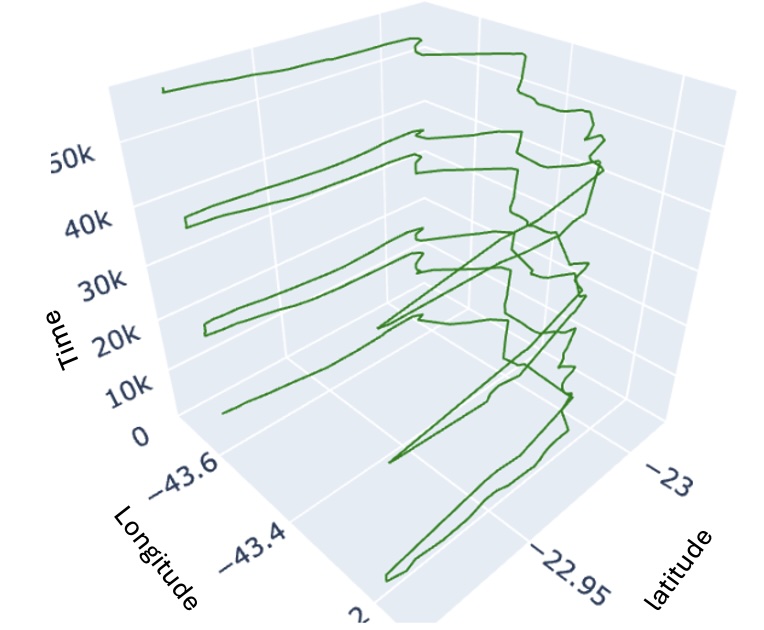}
    \includegraphics[width=.4\linewidth]{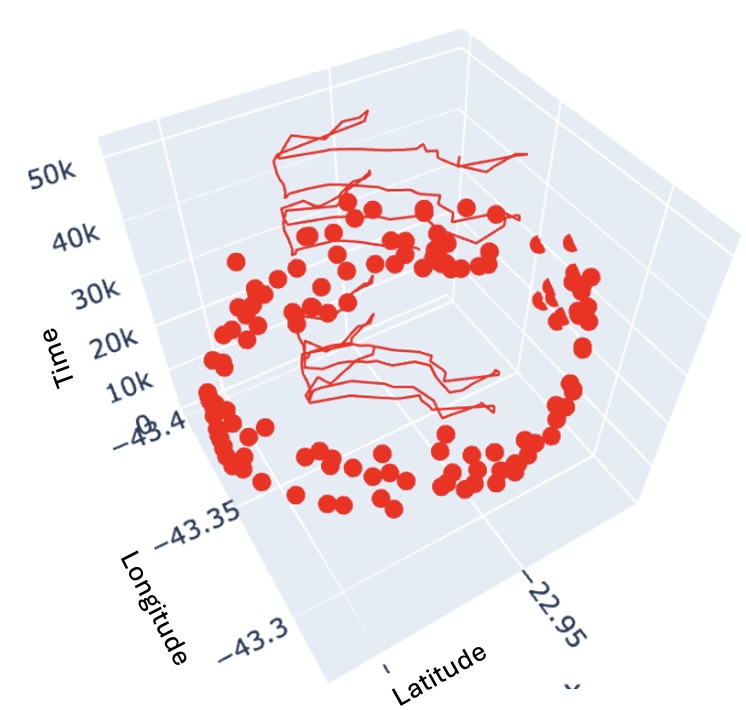}
    \caption{An unaugmented track (left) from the Rio Bus Dataset viewed in three dimensions and an augmented track (right) synthesized by inserting an anomaly in the middle third of the original track.}
    \label{fig:track_augmentation2}
\end{figure}

\begin{figure}[H]
    \centering
    \includegraphics[width=.8\linewidth]{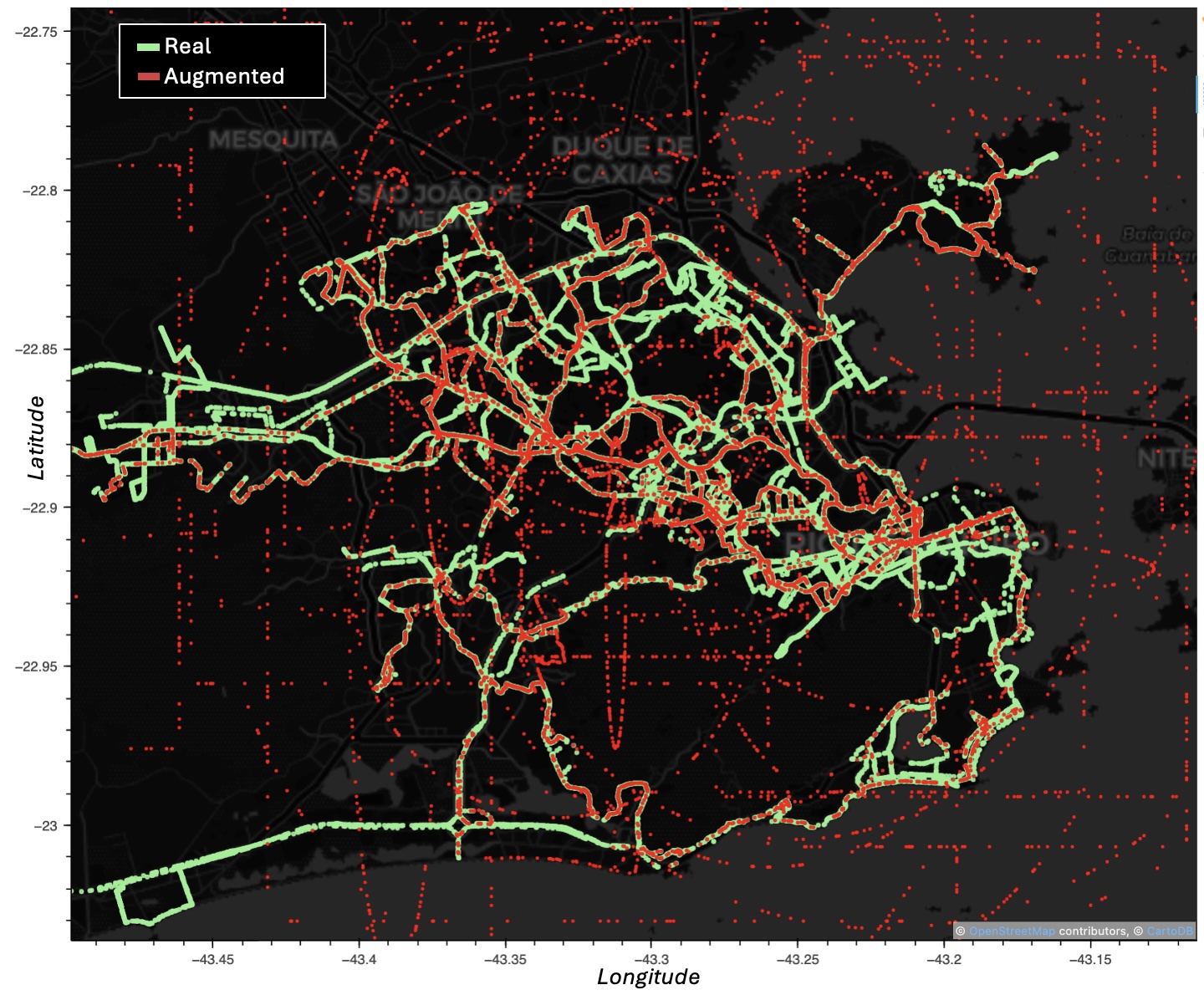}
    \caption{Synthetic data generation using the Rio Bus Dataset.}
    \label{fig:rio}
\end{figure}

Fig. \ref{fig:track_augmentation2} shows the methodology for generating synthetic data with crop circles similar to those found in \cite{Harris}.
We generate anomalies to include circles, squares, and ellipses of various sizes within the Rio Bus Dataset (see Fig. \ref{fig:rio}).
We ensure that the radius of the anomaly generated is no larger than 50\% of the original track length.

The resulting dataset is comprised of approximately 1,000 anomalous tracks and 3,400 unmodified tracks.
We will use this data to determine the space-time metric that maximizes the discriminatory power of persistent homology.

\subsection{Persistence Diagrams for Trajectory Data}
\begin{figure}[H]
    \centering
    \includegraphics[width=.9\linewidth]{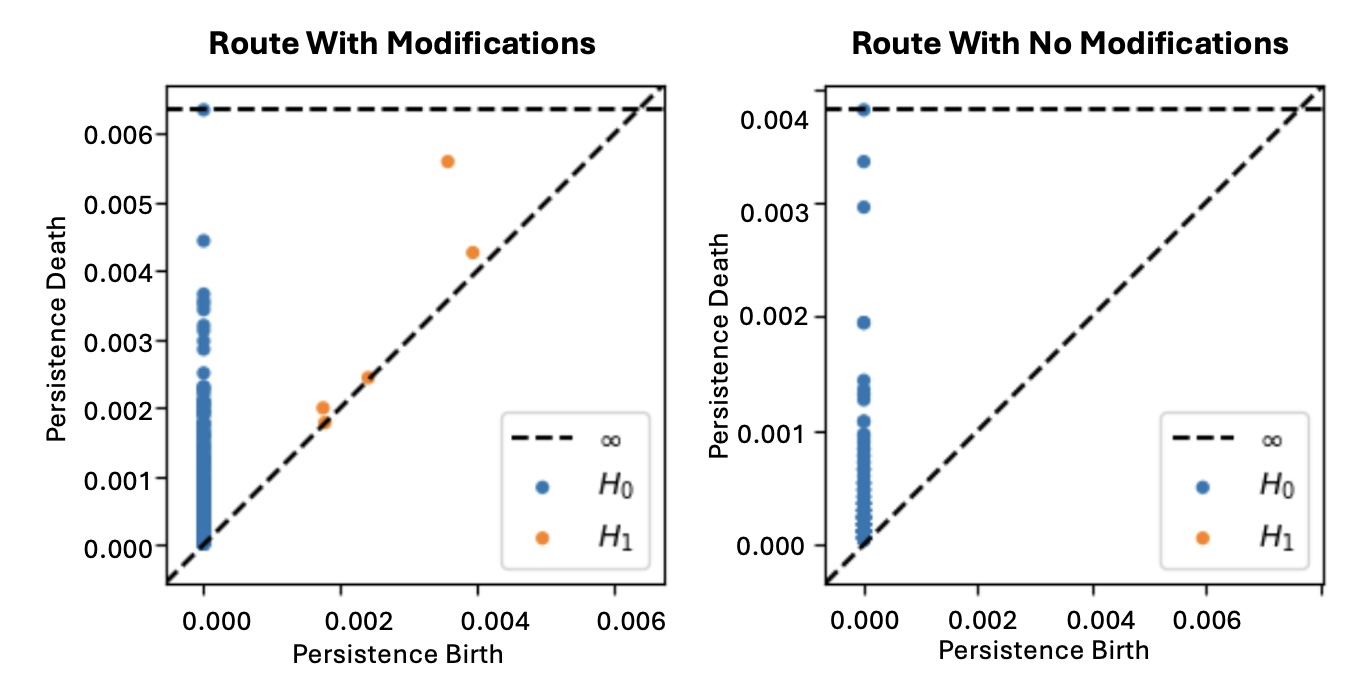}
    \caption{Examples of persistence  calculations for $0^{\text{th}}$ and $1^{\text{st}}$ homology generators. The modified route has significant generators in first homology not found in the unmodified route.}
    \label{fig:example_perst}
\end{figure}

For each trajectory, we generate a persistence diagram.
This calculation is done using \ttfamily giotto-tda\normalfont, a high-performance topological machine learning module in Python \cite{giotto}.
The example in Fig. \ref{fig:example_perst} shows the persistence diagrams generated from the metric with velocity parameter $k=10\text{ km/hr}$.

Fig. \ref{fig:example_perst} shows examples of the persistent homology calculations for two trajectories; one that has been modified and another that was unmodified.
Unsurprisingly, both examples have significant generators in the $0^{\text{th}}$ homology as they are connected components of the data.
The existence of significant generators in $1^{\text{st}}$ homology captures the loop structure found in the modified track.

\subsection{Features Extracted from Persistence Diagrams}
Once equipped with a persistence diagram, we identify meaningful features from the diagram that best summarize the loops in the data.
For instance, if there are multiple loops in the data, there may be multiple large generators in the persistence diagram.
Ideally, our chosen feature would capture the size and number of loops. 

There are many features that one could extract, including: sum of lifespans, Carlsson Coordinates, summary statistics, and entropy \cite{CC}.
These features can also be fine-tuned using noise-reduction techniques on the persistence diagram.
Other techniques have been used to determine the distance between persistence diagrams, such as using image processing and Wasserstein distance directly on the diagrams themselves \cite{giotto}.

\begin{figure}[H]
    \centering
    \includegraphics[width=.99\linewidth]{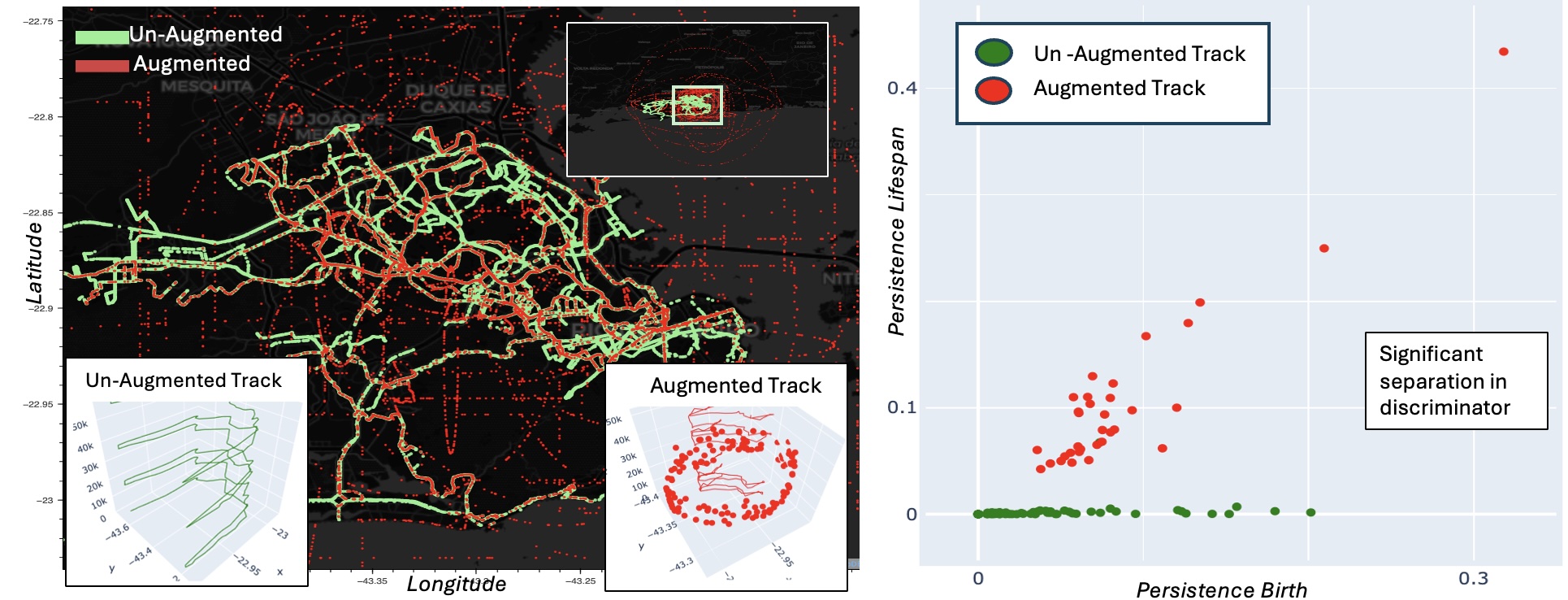}
    \caption{The discriminatory power of largest $1^{\text{st}}$ homology birth/death gap for augmented and un-augmented tracks}
    \label{fig:RioPH}
\end{figure}

While these approaches are useful in other contexts, it is sufficient here to consider simply the largest lifespan in the $1^{\text{st}}$ homology.
Therefore, the most prominent feature in the $d^{\text{th}}$ persistence homology is merely the longest lifespan among all features found in dimension $d$.
This is defined explicitly as
\[
m_d(\text{route}):=\max_{x_i}\{\text{death}(x_i)-\text{birth}(x_i)\},
\]
where $x_i$ is a generator of a feature in $d^{\text{th}}$ persistent homology for the route.
We are primarily interested in features extracted from the dimension-1 persistence diagrams, as generators of the $1^{\text{st}}$ homology are loops.
Fig. \ref{fig:RioPH} shows using persistence lifespan as a discriminator for anomalous and non-anomalous tracks in Rio Bus Dataset is effective. 

\subsection{Hyperparameter Tuning and Testing}
To select an appropriate velocity scale parameter $k$ for the distance metric, we perform a hyperparameter sweep over $k$.
This finds the optimal global transit rate that maximizes the gap in the maximum lifespan $m$ of a route with modifications versus that of a route without modifications.

\begin{figure}[H]
    \centering
    \includegraphics[width=.6\linewidth]{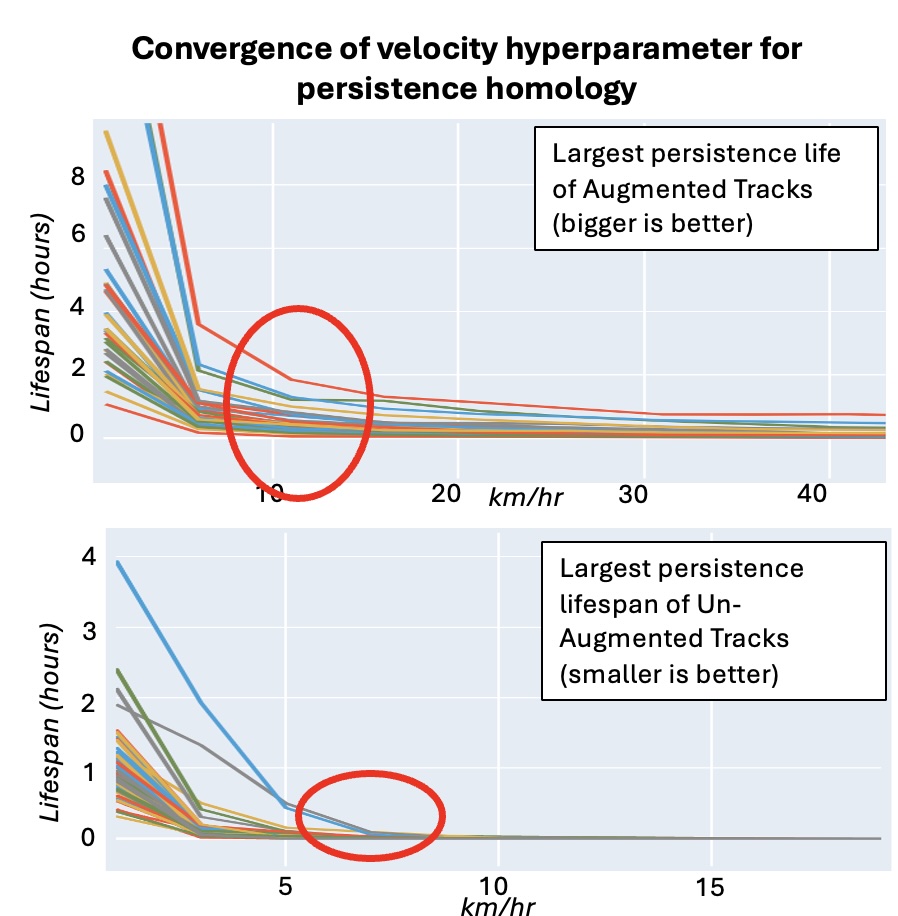}
    \caption{Hyperparameter convergence for $k$ km/hr. At approximately $10$ km/hr, we find persistent homology to be a good discriminator in Rio Bus Dataset for $k\geq10$.}
    \label{fig:ph}
\end{figure}

For each track in the Rio Bus Dataset, the persistent homology is computed from each metric defined for various values of $k$ (see Fig. \ref{fig:RioPH}).
The maximum lifespan among features detected in the persistence diagram of the $1^{\text{st}}$ homology is then computed, as shown in Fig. \ref{fig:ph}.
The optimal choice of $k$ is the one such that the plot values are large for the augmented tracks while small for the unaugmented tracks.
This empirically occurs near $k\approx10\text{ km/hr}$ as seen below. 

Fig. \ref{fig:ph} shows that the optimal choice of k appears to occur near $k\approx10\text{ km/hr}$, though values of $k$ between $10$ - $30\text{ km/hr}$ have roughly the same discriminatory power of classification.
To maximize resolution of feature identification, the value of $k$ is tuned to $10\text{ km/hr}$ for testing and is fixed at this value for the sequel.

\section{AIS Near San Francisco Example}
\begin{figure}[H]
    \centering
    \includegraphics[width=.7\linewidth]{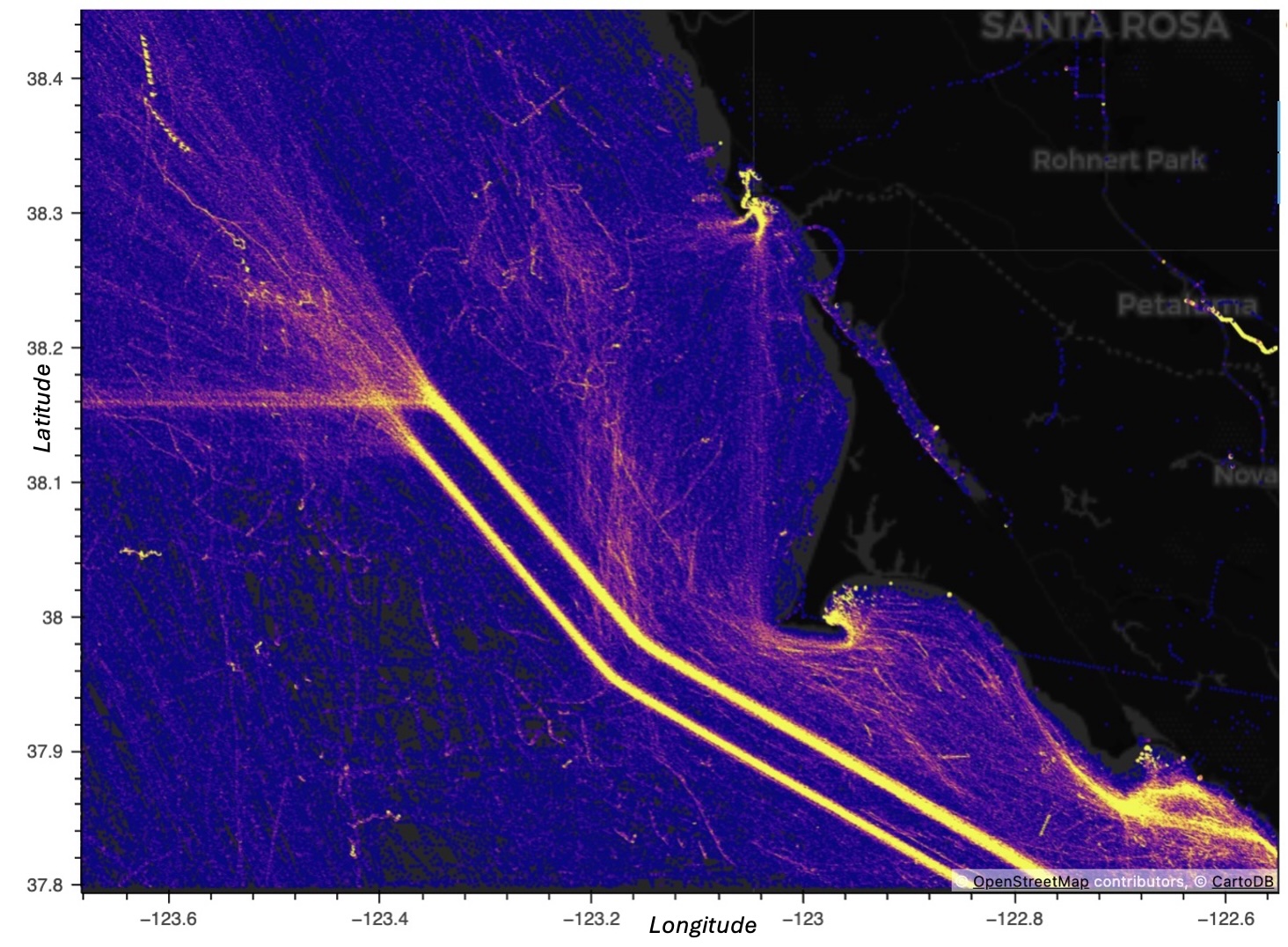}
    \caption{Tracks of around 1,500 MMSI's for testing algorithm spanning the dates from 2019/05/20 to 2020/04/05, the longitudes from $-123.3196^\circ$ to $-122.7372^\circ$, and the latitudes from $37.8406^\circ$ to $38.3473^\circ$ \protect\cite{marinecadastre}.}
    \label{fig:calidata}
\end{figure}

Using Automatic Identification System (AIS) data, a repository of global ship transponder data, we find examples of anomalous geolocations.
Within AIS data we consider only latitude, longitude, timestamp, and Maritime Mobile Service Identity (MMSI) data.
In this case, the MMSI is a unique identifier for each ship which we refer to as the selector.  
Following reports that anomalies were found near Point Reyes in \cite{skytruth}, we analyzed AIS data in the region to find new anomalies using our topological approach.
This data was provided by AccessAIS \cite{marinecadastre} for the extent listed in the caption of Fig. \ref{fig:calidata}.

Fig. \ref{fig:calidata} shows trajectory points highlighted by density for one year's worth of data near San Francisco.
We will next apply our detection method with hyperparameters set to the optimal tuning found in the previous section to identify anomalous trajectories.  

\subsection{Anomaly Identified From Persistence Diagrams of AIS Data}
\begin{figure}[H]
    \centering
    \includegraphics[width=.5\linewidth]{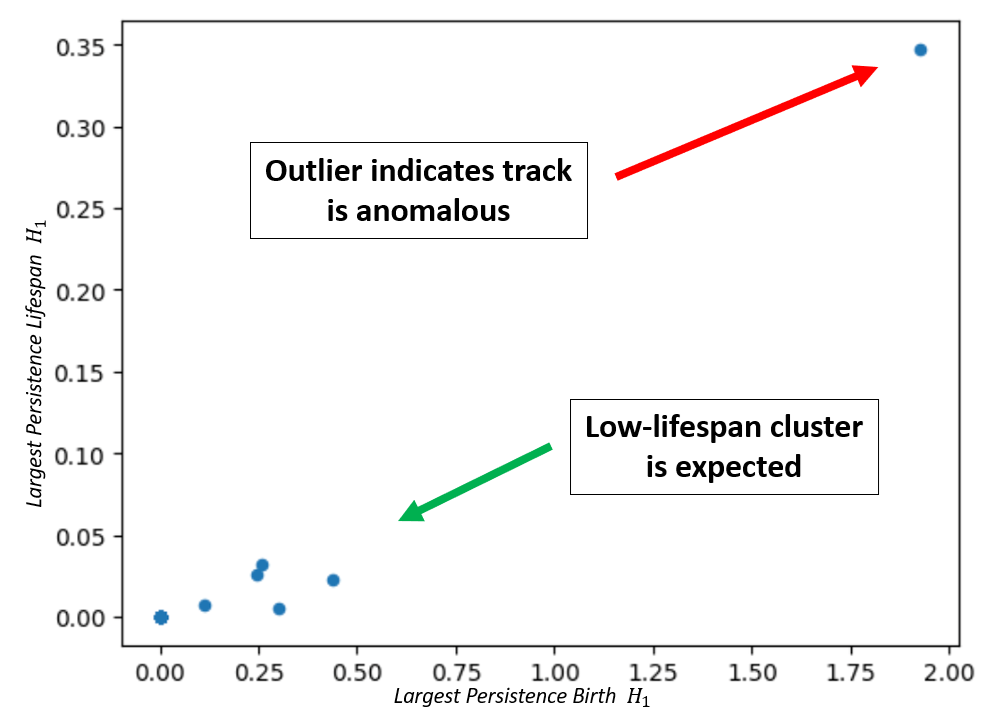}
    \caption{Plot of persistence lifespan vs. birth of AIS data off Point Reyes showing a significant outlier that corresponds to a previously unknown anomalous track.}
    \label{fig:pcali}
\end{figure}

For each MMSI within this dataset, we compute the persistent homology of the data points and extract the feature of longest lifespan in the $1^{\text{st}}$ homology.
Fig. \ref{fig:pcali} shows a plot of birth-versus-lifespan (noting that the plot's aspect ratio is not $1:1$ and that the main diagonal has been removed).
We see a clear outlier was found within the data.
Observe that most MMSI tracks are near $0$, as is expected.

\begin{figure}[H]
    \centering
    \includegraphics[width=.99\linewidth]{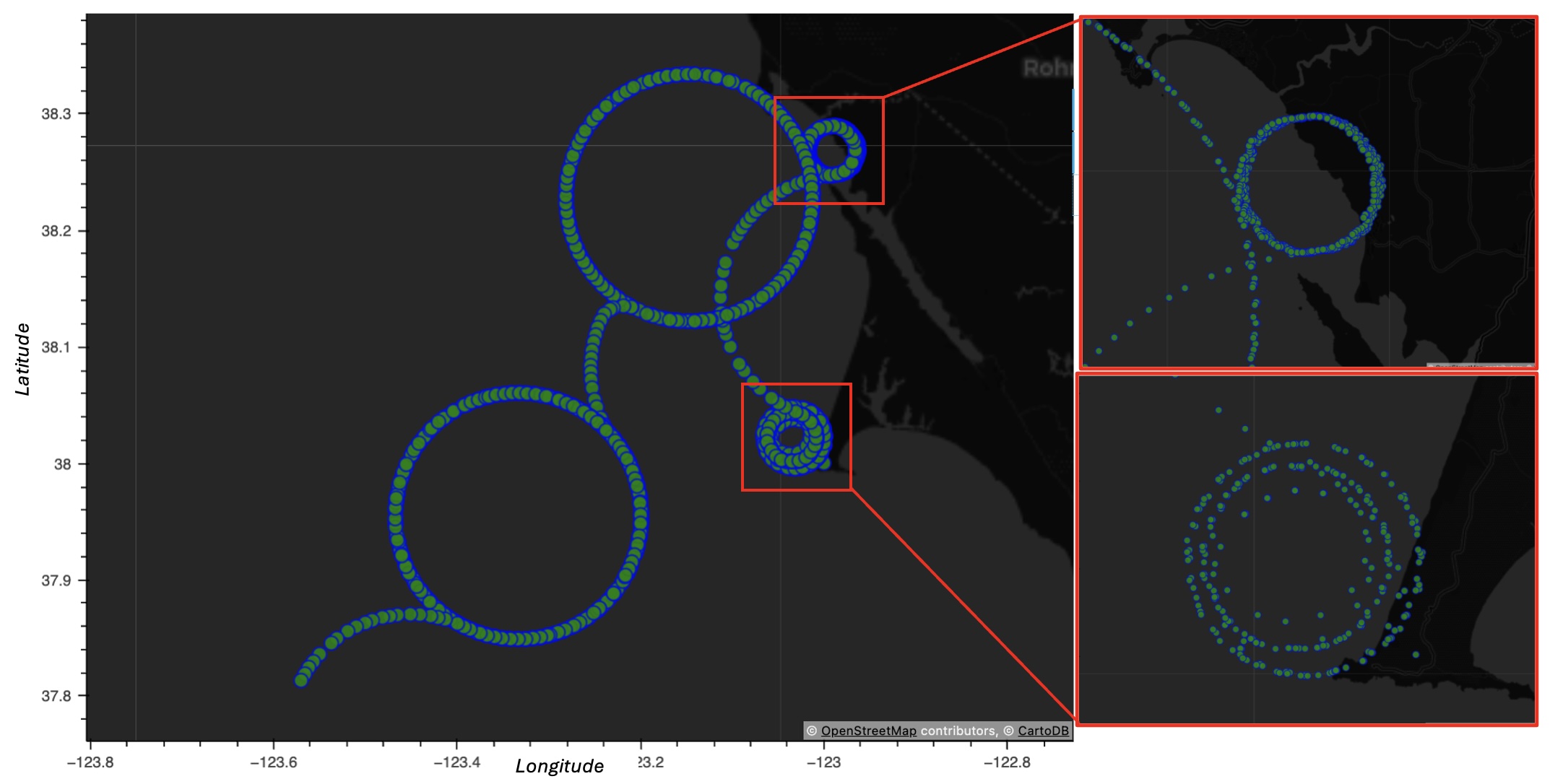}
    \caption{An anomalous trajectory of a cargo ship (MMSI 355692000) off Point Reyes between 2019/06/27 and 2019/06/29 identified using TDA-derived features.}
    \label{fig:PointReyesOutlier}
\end{figure}

Fig. \ref{fig:PointReyesOutlier} shows that the track corresponding to the outlier in the persistence diagram exhibits anomalous behavior; that is, it contains crop circles.
This outlier is shown to be anomalous by its crop circle pattern in addition to instances of it traveling over land in clearly unusual manner for a cargo ship.
To date, the authors do not know of any other instance of this cargo ship being identified as having anomalous geospatial  trajectories.

\section{Discussion}
This section discusses the results above and outlines considerations that should be addressed in any future lines of research.

\subsection{Pre-processing techniques necessary at scale}
The sampling rate found in AIS data is often quite high, which is beneficial in persistent homology calculations since this gives a finer resolution of the desired topological features.
Conversely, large gaps in data can easily lead to being unable to identify loop structures.
Such gaps in the data might be mitigated by pre-processing trajectories to include interpolation and would likely be beneficial to the overall analysis.

\subsection{Loops in Time}
It is possible for loops to exist entirely in a plane orthogonal to the $(\text{longitude},\text{latitude})$ plane.
In such a case, projecting these loops to that plane would result in a degenerate topological circle perceived as a curve.
This may be a rare case; however, the possibility exists and is worth noting since these loops may or may not also constitute a selector for identifying anomalous track data, and yet our approach would still be able to identify their existence.
Such loops may or may not be anomalous; but the criteria necessary for their existence is quite specific, and we have no examples to date.

\subsection{Alternate Embeddings}
It would be interesting to consider different embeddings of trajectory data by exploring the effects of larger embedding dimension and employing different strides and time delays (mentioned in Section 2) in the computation of Takens' embedding.
There is also merit to investigating different hyperparameter choices for these embeddings, including the information theoretic approach found in \ttfamily giotto-tda \normalfont \cite{giotto}.
In particular, it is not clear how anomalous loops in track data would manifest or be interpreted in Takens' embeddings of dimensions greater than 3 (e.g., choosing integer multiples of 3 for the stride).

\subsection{Dimension Considerations}
We have only used the first two homology groups $H_0$ and $H_1$ in our detection scheme.
The homology groups $H_n$ for $n\geq2$ should also be considered in further analysis.
Recall that these higher-dimensional homology groups will determine the existence of other topological features in the point cloud.
For instance, we infer from a point cloud with nontrivial $H_2$ in its persistent homology that the data are attracted to a closed surface lying in the embedding space.
These higher-dimensional homology data, hence, could contain distinct information from that contained in $H_0$ and $H_1$ alone.

To this point, just as $H_0$ could be used to construct a classifier for point clusters, $H_2$ could be used to further classify anomalous tracks.
Explicitly, it would be interesting to determine if there is a relationship between any loops found in the data via $H_1$ and any surfaces found in $H_2$.
Perhaps the loops are embedded in these surfaces in specific ways (e.g., meridional or longitudinal embeddings of loops in the surface of a torus) and could lead to further insights about the types of anomalies in track data one expects to find.

\subsection{Pre-processing on Persistence Diagrams}
In creating a persistence diagram, generators with very small lifespans are considered to be noise; however, when summing over all of these lifespans, they may contribute to false positives.
Pre-processing noise on the diagrams may improve identification.
Entropy and information theoretic approaches exist to remove the noise, and it may be beneficial to apply them \cite{embeddelay}.

\subsection{Extracting More Advanced Features From Persistent Homology}
There are many features that can be extracted from a persistence diagram including summary statistics, Carlsson Coordinates, as well as more advanced entropy based techniques \cite{CC}.
Additionally, there are distance metrics defined on persistence diagrams such as bottleneck distance \cite{bottleneck} that can be used to cluster diagrams together.
In doing so, additional features could be extracted as well as the potential of clustering different anomaly types together. 

\subsection{Improved Metrics or Quasi-Metrics}
Significant improvements to the simplicial complex could be made that would better represent real-world events.
Appealing to the local construction of the distance metric would greatly improve the accuracy of when simplices are added to the complex and could remove the need the need to perform a hyperparameter sweep.
This could be done by finding a best fit function of all the velocities taken over all spatial coordinates found within the dataset, and this may even be extended to be a function of time.
Additionally, we could consider a quasi-metric where, instead of finding a function, we compute the distance matrix between all points, filling it in a manner that accounts for different velocities between the measured points depending on the direction of travel.
This quasi-metric need not be symmetrical and would offer a unique extension that would be useful in this case.
The authors do not know of any prior examples of this kind of implementation. 

\section{Conclusion}
In conclusion, our paper presents a novel method for detecting anomalous geospatial trajectories within Automatic Identification System (AIS) data, specifically focusing on a class of anomalies known as crop circles.
By leveraging topological data analysis (TDA), we introduce a scalable approach that analyzes trajectory data using persistent homology.

We contribute to the existing body of work by proposing a fundamentally different approach, choosing to analyze individual trajectories instead of trajectory data at a population level.
Furthermore, we exploit the inherent structure of spatiotemporal data by embedding it into $\mathbb{R}^3$, enabling us to use persistent homology to identify loops within trajectories.
These loops serve as indicators of anomalous behavior since, as under normal conditions, trajectories embedded with respect to time should not contain loops.

Our approach offers several advantages over existing methods.
By analyzing trajectories individually (they are numerous but typically small), our approach is more scalable and CPU-parallelizable.
Additionally, we use loops as an obstruction to normality, effectively performing outlier detection.
Furthermore, our method is robust to small perturbations, variations in geometric shape, and local coordinate projections.

Through experiments using synthetic and real-world AIS data, we demonstrate the effectiveness of our approach in identifying anomalous trajectories, including those containing crop circles.
Our findings highlight the potential of TDA techniques in geospatial anomaly detection, paving the way for further research and applications in this domain.

Moving forward, our work opens avenues for exploring more advanced features extracted from persistent homology, refining pre-processing techniques, and investigating alternate embeddings.
Overall, our paper contributes to advancing the field of anomaly detection in geospatial data using topological methods and offers promising new directions for future research.

%
%
%



\newpage
\bibliographystyle{unsrt}

\bibliography{references}
\end{document}